\documentclass[a4paper,11pt]{article}
\pdfoutput=1

\usepackage{jheppub}
\usepackage[T1]{fontenc}
\usepackage{multirow} 
\usepackage{placeins}
\title{\boldmath Two-scalar-field $f(R)$ Thick Branes, Gravitational Resonances and Quasinormal Modes }

\author[a]{Xin-Yi Pan,}
\author[a,1]{Heng Guo,\note{Corresponding author.}}
\author[a]{Jing-Xin Gong,}
\author[a,b]{and Hong-Tao Jiang}

\affiliation[a]{School of Physics,
              Xidian University,
              Xi'an 710071, People's Republic of China}

\affiliation[b]{Shanghai GTA Semiconductor Co., Ltd.,
              Shanghai 200123, People's Republic of China}

\emailAdd{23009200092@stu.xidian.edu.cn}
\emailAdd{hguo@xidian.edu.cn}
\emailAdd{24201214709@stu.xidian.edu.cn}
\emailAdd{hongtao\_jiang@gtasemi.com.cn}

\abstract{
In this paper, we investigate thick brane worlds in $f(R)$ gravity supported by two-scalar-field. The two-scalar sector provides an analytical warped background with tunable energy-density splitting, allowing us to test whether a Bloch-type internal structure can generate long-lived tensor perturbations resonances in the physically admissible region. We impose the positivity of \(f_R\equiv df/dR\), the derivative of the gravitational Lagrangian with respect to the Ricci scalar, which plays the role of an effective gravitational coupling in \(f(R)\) gravity. This separates the smooth ghost-free branch from a singular branch where this effective coupling vanishes. In the ghost-free branch, neither the relative-probability spectrum nor the phase-shift transmission spectrum shows narrow real-axis resonant peaks. These real-axis diagnostics indicate that the internal brane structure alone does not produce long-lived tensor resonances in the ghost-free region. Sharp quasi-localization peaks appear only in the singular branch, where the vanishing effective coupling induces divergent structures in the tensor potential; these peaks should therefore be interpreted as singular-boundary signals rather than ghost-free resonances of the smooth brane background. We then characterize the ghost-free massive Kaluza-Klein modes in the complex-frequency plane. Using the Asymptotic Iteration Method where applicable and time-domain evolutions with a supersymmetric partner potential as a zero-mode filtering tool, we extract the fundamental quasinormal frequencies. The modes have negative imaginary parts and quality factors \(Q\simeq0.9-1.9\), showing that the ghost-free massive tensor excitations are broad, short-lived dissipative modes. Thus the QNM spectrum provides the appropriate complex-frequency description of the Kaluza-Klein dynamics when no narrow real-axis resonances are resolved.
}

\begin{document} 
\maketitle
\flushbottom

\section{Introduction}
\label{sec:intro}

The observation of gravitational waves (GWs) by the LIGO-Virgo-KAGRA
collaborations has opened a new observational window for testing gravity and
probing compact objects in the strong-field regime~\cite{LIGOScientific:2018mvr,
LIGOScientific:2020ibl, KAGRA:2021vkt, LIGOScientific:2025slb}. The growing
catalog of GW events has enabled increasingly precise tests of General
Relativity (GR)~\cite{LIGOScientific:2016lio, LIGOScientific:2018dkp,
LIGOScientific:2019fpa, LIGOScientific:2020tif, LIGOScientific:2021sio,
LIGOScientific:2025rid, LIGOScientific:2025obp}. The same observational
progress also motivates future cosmological probes with space-based GW networks
and standard sirens~\cite{Guo:2022sts}. During the ringdown phase of a perturbed
open system, the response is governed by quasinormal modes (QNMs). A QNM
frequency is complex: its real part gives the oscillation scale, while its
imaginary part determines the damping rate associated with energy leakage.
Although QNMs have been studied most extensively for black holes and black
branes~\cite{Kokkotas:1999bd, Nollert:1999ji, Berti:2009kk, Konoplya:2011qq,
Cardoso:2016rao, Jusufi:2020odz, Cheung:2021bol, Giesler:2019uxc,
Vishveshwara1970, Nollert:1996rf, Onozawa:1995vu, Andersson:1996xw,
Onozawa:1996ux, Leung:1997was, Zhang:2006hh, Miranda:2008vb, Horowitz:1999jd,
Konoplya:2003dd, Kanti:2001cj, Cardoso:2001bb, Toshmatov:2016bsb}, the same
spectral idea is useful in many open wave systems~\cite{Demesy:23,
PhysRevA.49.3057, PhysRevX.11.041020}. This makes QNMs a natural tool for
describing dissipative excitations in braneworld scenarios as well.

Braneworld models provide a framework in which our four-dimensional universe is
embedded in a higher-dimensional bulk. Such ideas are rooted in early attempts
to unify interactions through extra dimensions~\cite{Nordstrom, Kaluza:1921,
Klein:1926}, and were developed further in domain-wall models~\cite{Akama:1982jy,
Rubakov:1983bb}, the Arkani-Hamed-Dimopoulos-Dvali scenario~\cite{ArkaniHamed:1998rs,
Antoniadis:1998ig}, and the Randall-Sundrum models~\cite{Randall:1999ee,
Randall:1999vf}. In thin-brane constructions the brane is idealized as a
singular hypersurface. Thick branes, generated by scalar fields coupled to
gravity, replace this idealization with a smooth domain-wall-like geometry and
allow one to study field localization and perturbation dynamics in a regular
background~\cite{DeWolfe:1999cp, Gremm:1999pj, Csaki:2000fc, Campos:2001pr,
Bazeia:2003aw, Dzhunushaliev:2010fqo, Dzhunushaliev:2011mm,
Kobayashi:2001jd, Kehagias:2000au, Melfo:2002, Liu:2007ku, Liu:2007gk}.
Over the past decades, braneworld models have also been applied in a range of
contexts, including effective four-dimensional gravity, field localization,
black holes, cosmology, and holography~\cite{Shiromizu:1999wj, Tanaka:2002rb,
Gregory:2008rf, Jaman:2018ucm, Adhikari:2020xcg, Geng:2020fxl, Geng:2021iyq,
Geng:2022dua, Bhattacharya:2021jrn, Dzhunushaliev:2009va, Maartens:2010ar,
Herrera:2010, Liu:2017gcn, Guo:2023mki, Ahluwalia:2022ttu}.

The structure of a thick brane is determined jointly by the matter sector and
the gravitational dynamics. In single-scalar models, the same scalar degree of
freedom is usually tied to the formation of the smooth defect, the support of
the warped geometry, and the shape of the internal energy-density profile.
Multi-field scalar sectors provide a more flexible possibility: different
directions in field space may play different dynamical roles, and their mutual
interaction can generate internal structures that are absent in minimal
single-field realizations.

This idea appears in different contexts. In cosmology, two-scalar-field models
have been used to build unified descriptions of inflation, dark matter, and
dark energy, where the two fields are assigned different roles at different
cosmic epochs and the resulting background evolution can leave imprints in the
gravitational-wave spectrum~\cite{Sa:2020triple, Luongo:2026twoScalarGW}.
Although our setup is a five-dimensional brane model rather than a
four-dimensional cosmological model, the same general lesson is useful:
a two-scalar sector provides a controlled field-space structure for realizing
richer background dynamics and for testing how such structures are encoded in
gravitational perturbations.

In the braneworld context, the standard example is the Bloch brane, where the
interaction of two scalar fields leads to a splitting of the brane core and a
double-peak energy-density distribution~\cite{Bazeia:2004wc, Almeida:2009jc,
Cruz:2013uwa, Fu:2011pu}. Related two-field constructions, such as hybrid Bloch
branes, further show that the scalar interaction can control the internal
energy-density profile and modify the gravitational volcano potential
experienced by perturbations~\cite{Bazeia:2017hybrid}. Thus, the value of a
two-scalar sector is not merely technical. It provides an analytical and tunable
matter source with which one can test whether internal brane structure has a
genuine spectral consequence.

In parallel, modified gravity theories provide additional ways to change the
background geometry and the perturbation potential. Among these, \(f(R)\)
gravity, in which the Einstein-Hilbert Lagrangian is replaced by a function of
the Ricci scalar \(R\), is one of the most studied extensions of
GR~\cite{Sotiriou:2010, DeFelice:2010, Nojiri:2007, Capozziello:2005,
Starobinsky:2007, Hu:2007}. Other modified-gravity frameworks, including
teleparallel models, have also been widely explored~\cite{Hayashi:1979wj,
Aldrovandi:2013wha, Ferraro:2006jd, Bengochea:2008gz, Linder:2010py,
Karami:2010bys, Cai:2015emx, Krssak:2018ywd, Bahamonde:2021gfp}. Thick branes
in modified gravity have therefore attracted considerable
attention~\cite{Afonso:2007gc, Liu:2011wi, Bazeia:2013uva, Zhong:2016iko,
Zhong:2017ffr, Balcerzak:2010kr, Bazeia:2014poa, Guo:2015, Tan:2020sys,
Geng:2015kvs, Gu:2016nyo, Guo:2019vvm, Guo:2025geo, Xie:2021ayr,
Chen:2020zzs, Moreira:2021uod, Silva:2022pfd, Xu:2022xxd}.

Thick branes in \(f(R)\) gravity are technically nontrivial because the field
equations are generally of fourth order. Exact analytical solutions are
therefore difficult to obtain without additional assumptions or extra degrees
of freedom. For the quadratic model \(f(R)=R+\alpha R^2\), where \(\alpha\)
denotes the strength of the quadratic curvature correction, analytical
thick-brane backgrounds with a tunable thickness have been constructed and
analyzed in detail~\cite{Bazeia:2013uva, Xu:2015}. In these backgrounds, the
curvature correction modifies the warped geometry and can change the
energy-density profile of the brane.

The two-scalar realization used in this work should be understood as a concrete
implementation of the above multi-field idea in a warped extra dimension. The
two scalar profiles provide an explicit analytical matter source whose combined
energy density can interpolate between a single-core thick brane and a
Bloch-type split brane. This gives a controlled setting in which the internal
structure of the brane can be varied while the tensor perturbation problem
remains tractable. At the same time, the tensor perturbation equation does not
couple directly to the individual scalar components. Once the background
geometry and the quantity \(f_R\equiv df/dR\), which acts as an effective
gravitational coupling in the \(f(R)\) sector, are fixed, the tensor spectrum is
governed by these background functions. The central question is therefore
whether the internal structure realized by the two-scalar source can leave a
genuine spectral imprint in the physically admissible ghost-free region.

For multi-scalar branes, the scalar perturbation sector can be considerably
more involved than in single-scalar models, since perturbations along different
field-space directions may be coupled~\cite{Chen:2018multiScalar}. In the
present work we therefore focus on the transverse-traceless tensor sector,
where the gravitational spectral problem is cleanly defined. In this sector,
tensor perturbations obey a Schr\"{o}dinger-like equation with an effective
potential determined by the warped geometry and by the effective coupling
\(f_R\)~\cite{Zhong:2011, Zhong:2016iko, Zhong:2017ffr}. The massless graviton
zero mode must be localized in order to reproduce four-dimensional Newtonian
gravity, while the massive Kaluza-Klein (KK) modes form a continuum in models
with an infinite extra dimension. These massive modes may show
quasi-localization near the brane, appearing as resonant structures in
real-axis diagnostics such as the relative probability, transmission
coefficient, or phase shift~\cite{Liu:2009ve, Almeida:2009jc, Cruz:2013uwa,
Clarkson:2005mg, Seahra:2005wk, Seahra:2005iq, Tan:2022uex}. They may also
appear as QNMs, namely poles in the complex frequency plane with finite damping
rates~\cite{Tan:2022vfe, Tan:2023cra, Tan:2024url, Jia:2024sdk, Jia:2024pdk,
Zhu:2024gvl, Tan:2024qij, Deng:2025}. These two descriptions are
complementary: a long-lived resonance close to the real axis can produce a
sharp real-axis peak, whereas a short-lived dissipative excitation may leave no
narrow peak and is more naturally characterized by its complex QNM frequency.

This distinction is central to the present work. Previous studies of
\(f(R)\)-brane structures have shown that an internal brane structure does not
automatically imply the existence of graviton resonances, and that sharp
real-axis peaks may instead be associated with singular structures of the
effective potential~\cite{Xu:2015}. Recent studies of QNMs in \(f(R)\) thick
branes have further shown that complex-frequency methods are useful for
characterizing dissipative KK excitations~\cite{E:2026}. Motivated by these
developments, we ask a more specific question: does the Bloch-type internal
structure of a two-scalar \(f(R)\) brane generate long-lived tensor resonances
in the ghost-free region, or are the physically admissible massive modes better
described as short-lived QNMs?

We answer this question by performing a real-axis and complex-frequency spectral
analysis of the two-scalar \(f(R)\) thick brane. First, we determine the
physically relevant parameter regions by imposing both the scalar-field reality
condition and the ghost-free condition, namely the positivity of the effective
coupling \(f_R\). This separates the smooth ghost-free branch from a singular
branch where this effective coupling vanishes. Second, we use the
relative-probability method as a real-axis diagnostic for narrow quasi-localized
tensor modes. In the ghost-free branch, no sharp peaks are resolved. To check
that this result is not merely an artifact of the relative-probability
criterion, we also compute the even and odd scattering phase shifts and the
associated transmission coefficient. The transmission spectrum increases
smoothly with the KK mass and shows no narrow resonant peaks. These results
indicate that the smooth ghost-free potentials do not support long-lived tensor
resonances detectable on the real mass axis.

The singular branch behaves differently. For sufficiently thick branes and
positive curvature correction, the effective coupling \(f_R\) can vanish at
finite positions in the extra dimension, and the tensor effective potential
then develops singular structures through inverse powers of \(f_R\). In this
branch, sharp relative-probability peaks appear, and the corresponding wave
functions are strongly enhanced near the brane. Since this region violates the
ghost-free condition, these peaks should be interpreted as singular-boundary
quasi-localization signals rather than resonances of the smooth ghost-free
brane background. This separation provides a useful diagnostic: the tensor
spectrum distinguishes short-lived dissipative KK modes in the ghost-free
branch from singular-boundary quasi-localization caused by the vanishing
effective coupling.

To describe the massive KK excitations in the ghost-free branch, we extend the
analysis to the complex frequency plane. We compute QNFs using the Asymptotic
Iteration Method where applicable, and we also extract them from time-domain
ringdown waveforms~\cite{Cooper:1994eh, Ge:2018vjq, Ciftci:2003As,
ciftci:2005co, AIM_2011, Ciftci:2005xn, Bayrak:2006qt, Champion:2008hg,
BSM_2023, DIM_2013, Megevand:2007uy, Press1971, Davis1971, Cunningham1978,
Detweiler1977, Detweiler1979}. In the time-domain calculation, the
supersymmetric partner potential is used as a practical tool to suppress the
non-dissipative graviton zero-mode contribution and to isolate the massive
ringdown signal. The extracted modes have negative imaginary parts, indicating
damped tensor perturbations. Moreover, their quality factors are of order
unity, \(Q=\omega_R/(2|\omega_I|)\sim O(1)\), which means that the amplitudes
decay within only a few oscillation periods. This quantitatively explains why
the corresponding modes do not appear as narrow peaks in the real-axis spectra:
they are broad, short-lived dissipative KK excitations rather than long-lived
quasi-localized resonances.

The structure of this paper is organized as follows. In
section~\ref{sec:background}, we present the two-scalar \(f(R)\) thick-brane
background, discuss the energy-density splitting, and identify the ghost-free
and singular regions of parameter space. In section~\ref{sec:perturbation}, we
derive the tensor perturbation equation, construct the supersymmetric partner
potential, and analyze the real-axis spectra using the relative probability and
phase-shift transmission diagnostics. In section~\ref{sec:qnm}, we study the
QNMs by combining the AIM with time-domain numerical evolution and quantify the
short-lived nature of the ghost-free massive KK modes. Section~\ref{sec:conclusion}
summarizes the main physical conclusions and discusses possible extensions.

\section{The two-scalar \texorpdfstring{$f(R)$}{f(R)} thick brane model}
\label{sec:background}

We consider a five-dimensional bulk spacetime in which \(f(R)\) modified gravity is minimally coupled to two real scalar fields \(\phi_1\) and \(\phi_2\)~\cite{Bazeia:2004wc, Afonso:2007gc, Bazeia:2013uva}. The total action is
\begin{equation}
S = \int d^5x \sqrt{-g} \left[ \frac{1}{2\kappa_5^2} f(R) + \mathcal{L}_m \right] \,,
\label{eq:action}
\end{equation}
where \(g\) is the determinant of the five-dimensional metric tensor \(g_{MN}\). The five-dimensional gravitational coupling constant is denoted by \(\kappa_5^2 = 8\pi G_5\), and we set \(\kappa_5 = 1\) in the following. The gravity sector is specified by
\begin{equation}
f(R) = R + \alpha R^2 \,,
\label{eq:fR_form}
\end{equation}
where \(R\) is the five-dimensional Ricci scalar and \(\alpha\) controls the quadratic curvature correction~\cite{Sotiriou:2010, DeFelice:2010}. The matter Lagrangian for the two background scalar fields is
\begin{equation}
\mathcal{L}_m = - \frac{1}{2} \partial_M \phi_1 \partial^M \phi_1
- \frac{1}{2} \partial_M \phi_2 \partial^M \phi_2
- V(\phi_1, \phi_2) \,,
\label{eq:matter_lagrangian}
\end{equation}
where \(V(\phi_1,\phi_2)\) is the scalar interaction potential, and the capital Latin indices \(M,N=0,1,2,3,5\) label the five-dimensional bulk coordinates~\cite{DeWolfe:1999cp}.

Varying the action~\eqref{eq:action} with respect to \(g_{MN}\) and \(\phi_i\) gives
\begin{equation}
f_R R_{MN} - \frac{1}{2} f(R) g_{MN}
- \left(\nabla_M \nabla_N - g_{MN} \Box^{(5)}\right) f_R
= T_{MN} \,,
\label{eq:grav_eq}
\end{equation}
\begin{equation}
\Box^{(5)} \phi_i - \frac{\partial V}{\partial \phi_i} = 0 \,,
\quad (i=1,2).
\label{eq:scalar_eq}
\end{equation}
Here \(R_{MN}\) is the five-dimensional Ricci tensor, \(\nabla_M\) denotes the covariant derivative, and \(\Box^{(5)} \equiv g^{MN}\nabla_M\nabla_N\). The quantity
\[
f_R \equiv \frac{df(R)}{dR}=1+2\alpha R
\]
controls the effective gravitational coupling in the \(f(R)\) theory. The energy-momentum tensor of the scalar fields is
\begin{equation}
T_{MN} = \sum_{i=1}^2 \partial_M \phi_i \partial_N \phi_i
+ g_{MN} \mathcal{L}_m \,.
\label{eq:energy_momentum}
\end{equation}

We focus on a static flat brane described by
\begin{equation}
ds^2 = g_{MN} dx^M dx^N
= e^{2A(y)} \eta_{\mu\nu} dx^\mu dx^\nu + dy^2 \,,
\label{eq:metric}
\end{equation}
where \(y\) is the extra-dimensional coordinate and \(e^{2A(y)}\) is the warp factor. The four-dimensional coordinates are denoted by \(x^\mu\) with \(\mu,\nu=0,1,2,3\), and \(\eta_{\mu\nu}=\mathrm{diag}(-1,1,1,1)\) is the Minkowski metric~\cite{Randall:1999vf, Csaki:2000fc}.

Assuming that the scalar fields depend only on the extra dimension, \(\phi_i=\phi_i(y)\), the relevant background equations can be written as~\cite{Zhong:2011, Xu:2015}
\begin{equation}
\phi_1'' + 4A' \phi_1' = \frac{\partial V}{\partial \phi_1} \,,
\label{eq:eom1}
\end{equation}
\begin{equation}
\phi_2'' + 4A' \phi_2' = \frac{\partial V}{\partial \phi_2} \,,
\label{eq:eom2}
\end{equation}
\begin{equation}
(\phi_1')^2 + (\phi_2')^2
= -f_R'' + A' f_R' - 3 f_R A'' \,,
\label{eq:eom3}
\end{equation}
\begin{equation}
V(\phi_1, \phi_2)
= -\frac{1}{2}
\left[
f_R'' + 7A' f_R'
- (5A'' + 8A'^2) f_R
- f(R)
\right] \,,
\label{eq:eom4}
\end{equation}
where the prime denotes the derivative with respect to \(y\).

To construct analytical backgrounds, we take the warp factor ansatz
\begin{equation}
e^{A(y)} = \operatorname{sech}^B(ky) \,,
\label{eq:warp_ansatz}
\end{equation}
where \(B>0\) controls the brane thickness and \(k\) sets the inverse length scale. In the numerical spectral analysis below we set \(k=1\), following the convention used in related studies~\cite{Xu:2015, Tan:2023cra, E:2026}. With this choice, one convenient analytical realization of the two scalar profiles is
\begin{equation}
\phi_1(y)
= \pm \sqrt{\frac{3}{2}B + 4B \alpha (16B + 4)} \, \tanh(y) \,,
\label{eq:phi1_sol}
\end{equation}
\begin{equation}
\phi_2(y)
= \pm \sqrt{\frac{3}{2}B - 4B \alpha (5B^2 + 16B + 8)}
\, \operatorname{sech}(y) \,.
\label{eq:phi2_sol}
\end{equation}
These fields provide an explicit matter source supporting the chosen warped geometry. The important point for the perturbation analysis is that the tensor sector will be governed by the resulting functions \(A(y)\) and \(f_R(y)\), rather than by a direct coupling to the individual scalar components.

\begin{figure}[tbp]
\centering
\includegraphics[width=0.95\textwidth]{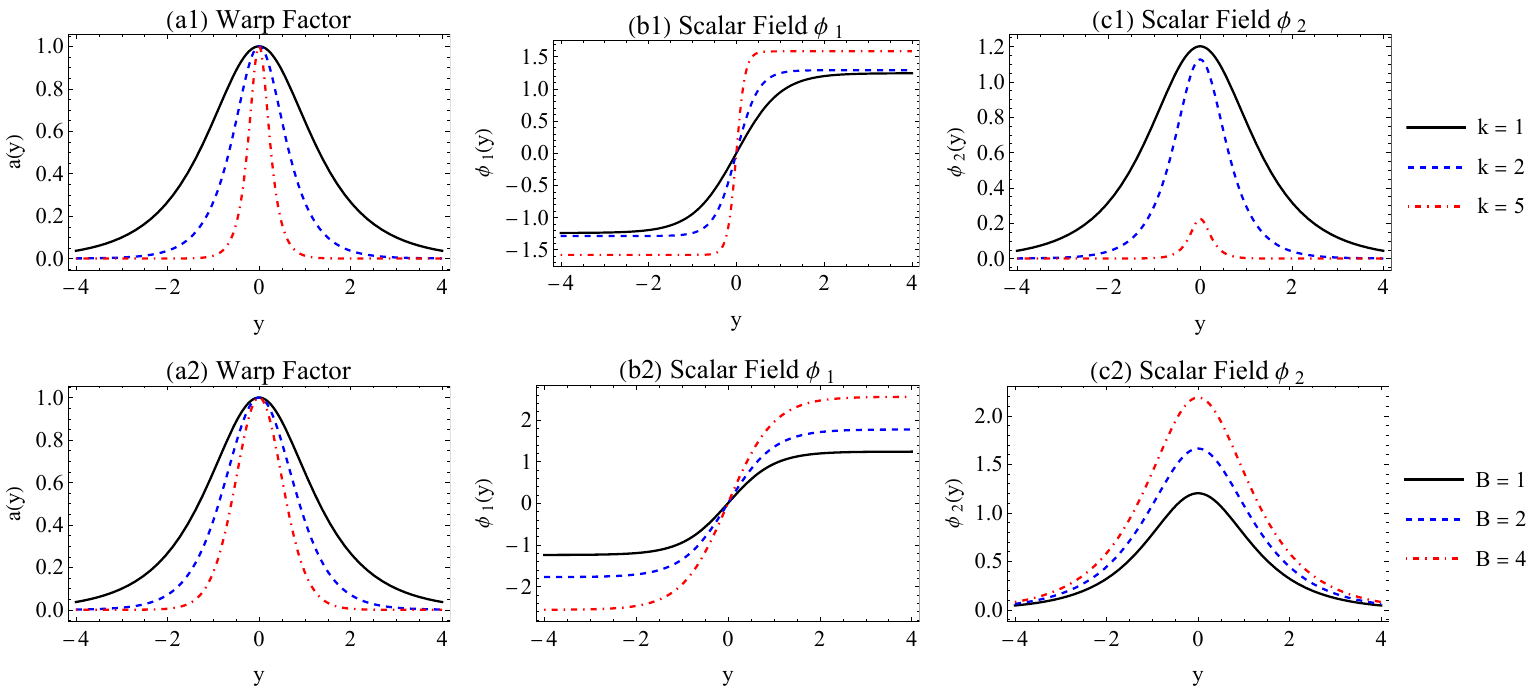}
\caption{\label{fig:profiles}
Profiles of the warp factor \(a(y)=e^{A(y)}\) and the background scalar fields \(\phi_1(y)\), \(\phi_2(y)\). The upper row illustrates the dependence on the scale parameter \(k\), while the lower row shows the dependence on the brane-thickness parameter \(B\).}
\end{figure}

The background profiles are shown in figure~\ref{fig:profiles}. The upper panels illustrate the effect of varying the scale parameter \(k\), while the lower panels show the dependence on the brane-thickness parameter \(B\). Increasing \(B\) makes the warp factor decay more rapidly away from the brane center. The geometry therefore approaches a thinner and more localized configuration, and the asymptotic bulk curvature becomes larger in magnitude, with the effective AdS cosmological constant scaling as
\[
\Lambda_{\rm eff}=-4B^2k^2 \qquad \text{\cite{Xu:2015}} .
\]
The scalar amplitudes change accordingly, reflecting the adjustment of the matter source required to support a steeper warped background.

As will be shown in section~\ref{sec:perturbation}, the same geometric compression also affects the tensor effective potential. Larger \(B\) generally makes the potential more localized and changes the transparency of the extra-dimensional scattering problem. This provides one of the channels through which the background geometry influences the massive Kaluza-Klein spectrum~\cite{Tan:2023cra, E:2026, Seahra:2005wk, Liu:2009ve}.

The internal structure of the brane can be characterized by the energy density distribution \(\rho(y)\equiv T_{00}\). For the present two-scalar background,
\begin{equation}
\rho(y)
= e^{2A(y)}
\left[
\frac{1}{2}(\phi_1'(y))^2
+ \frac{1}{2}(\phi_2'(y))^2
+ V(\phi_1,\phi_2)
\right] \,.
\label{eq:energy_density_formula}
\end{equation}
Substituting the background solution gives~\cite{Xu:2015}
\begin{equation}
\begin{aligned}
\rho(y) = B \, \operatorname{sech}^{2B}(y) \bigg\{&
-3B + 3\left(B+\frac{1}{2}\right)\operatorname{sech}^2(y) \\
&+ 4\alpha
\Big[
5B^3
- (10B^3+37B^2+32B+8)\operatorname{sech}^2(y) \\
&\hspace{2.0cm}
+ (5B^3+37B^2+44B+12)\operatorname{sech}^4(y)
\Big]
\bigg\} \,.
\end{aligned}
\label{eq:energy_density_explicit}
\end{equation}

The parameters \(B\) and \(\alpha\) control different aspects of the background. The parameter \(B\) mainly determines the geometric width of the brane: larger \(B\) produces a narrower and steeper energy-density profile~\cite{Campos:2001pr}. The curvature parameter \(\alpha\), on the other hand, changes the relative contribution of the higher-curvature sector and can modify the internal distribution of the energy density.

\begin{figure}[tbp]
\centering
\includegraphics[width=1\textwidth]{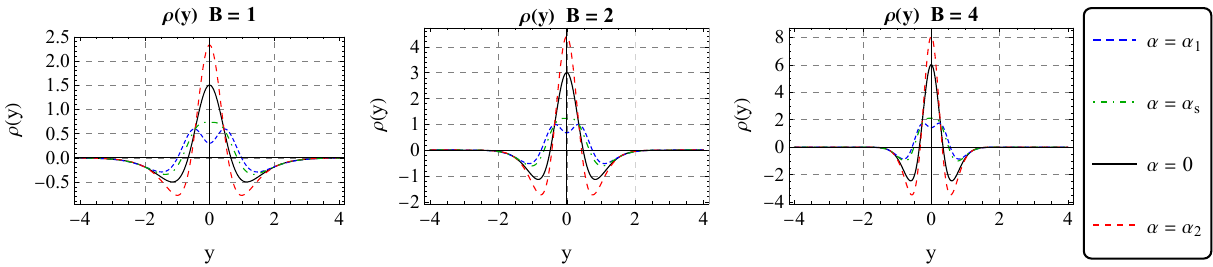}
\caption{\label{fig:energy}
Profiles of the energy density \(\rho(y)\) for \(B=1,2,4\). For each \(B\), four representative values of \(\alpha\) are shown: the GR limit \(\alpha=0\), the critical splitting point \(\alpha_s\), and the scalar-field reality bounds \(\alpha_1,\alpha_2\). The plots display the transition from a single-peak profile to a double-peak internal structure as \(\alpha\) reaches and passes below \(\alpha_s\).}
\end{figure}

For \(\alpha>\alpha_s\), the energy density has a single central peak. When
\[
\alpha \leq \alpha_s,
\qquad
\alpha_s \equiv
-\frac{3+9B}{8(16+60B+49B^2)} \,,
\]
the second derivative of \(\rho(y)\) at \(y=0\) changes sign and the central peak splits into two. This provides a Bloch-brane-like internal structure, as shown in figure~\ref{fig:energy}~\cite{Bazeia:2004wc, Almeida:2009jc, Cruz:2013uwa,Fu:2011pu}. In the following sections we will use this internally structured background to examine whether the matter-sector splitting leads to narrow tensor resonances in the physically admissible region.

The parameter space is constrained by two requirements. First, the scalar profiles must be real. Equivalently, the kinetic contribution in eq.~\eqref{eq:eom3} must be non-negative, which gives
\begin{equation}
\alpha_1 \equiv -\frac{3}{32(1+4B)} \,,
\qquad
\alpha_2 \equiv \frac{3}{8(8+16B+5B^2)} \,.
\label{eq:alpha_bounds}
\end{equation}
Thus the scalar-field reality condition restricts \(\alpha\) to the interval \([\alpha_1,\alpha_2]\).

Second, the ghost-free condition
\[
f_R=1+2\alpha R(y)>0
\]
must hold throughout the extra dimension. This condition is required to avoid the wrong-sign kinetic degree of freedom in the \(f(R)\) sector~\cite{Sotiriou:2010, DeFelice:2010, Zhong:2016iko}. For the present background,
\[
R(y)=-20B^2+(20B^2+8B)\operatorname{sech}^2(y),
\]
which decreases monotonically from \(R(0)=8B\) to \(R(\pm\infty)=-20B^2\). For \(\alpha\leq0\), the minimum of \(f_R\) occurs at the brane center,
\[
f_R(0)=1+16\alpha B.
\]
Within the scalar-field reality interval, this lower bound is satisfied. For \(\alpha>0\), the minimum of \(f_R\) occurs asymptotically,
\[
f_R(\pm\infty)=1-40\alpha B^2.
\]
Therefore the strict ghost-free condition imposes
\begin{equation}
\alpha < \alpha_k,
\qquad
\alpha_k \equiv \frac{1}{40B^2}.
\label{eq:alpha_k}
\end{equation}

Comparing \(\alpha_k\) with \(\alpha_2\), one finds that \(\alpha_k<\alpha_2\) for \(B>2\)~\cite{Xu:2015}. Hence, for \(B>2\), the positive-\(\alpha\) part of the physically admissible ghost-free region is cut off at \(\alpha_k\), not at \(\alpha_2\). For \(B<2\), the scalar-field reality bound is more restrictive. At \(B=2\), one has \(\alpha_k=\alpha_2\), so \(\alpha=\alpha_2\) should be understood as a limiting boundary value of the ghost-free region rather than as an interior point.

For \(B>2\), the interval
\[
\alpha_k \leq \alpha \leq \alpha_2
\]
belongs to a singular branch. In this branch \(f_R\) crosses zero at finite positions in the extra dimension. As will be seen explicitly in section~\ref{sec:perturbation}, the tensor effective potential contains inverse powers of \(f_R\), such as \(f_R^{-1}\) and \(f_R^{-2}\). The zero of \(f_R\) therefore produces singular structures in the effective potential~\cite{Balcerzak:2010kr, Bazeia:2014poa}. Since this branch violates the ghost-free condition, any sharp real-axis quasi-localization peaks appearing there should be interpreted as signals of the singular boundary rather than as ghost-free resonances of the smooth brane background.

This separation of parameter space is essential for the spectral analysis below. In the smooth ghost-free region, the effective potential defines an open scattering problem for the massive tensor modes. As shown later, this region does not support narrow real-axis tensor resonances detectable by the relative-probability or transmission diagnostics; instead, the corresponding massive KK excitations are described by short-lived QNMs in the complex frequency plane. In the singular branch, by contrast, the potential can develop sharp structures and produce pronounced real-axis quasi-localization peaks. These two cases should be physically distinguished.

\begin{figure}[tbp]
\centering
\includegraphics[width=0.7\textwidth]{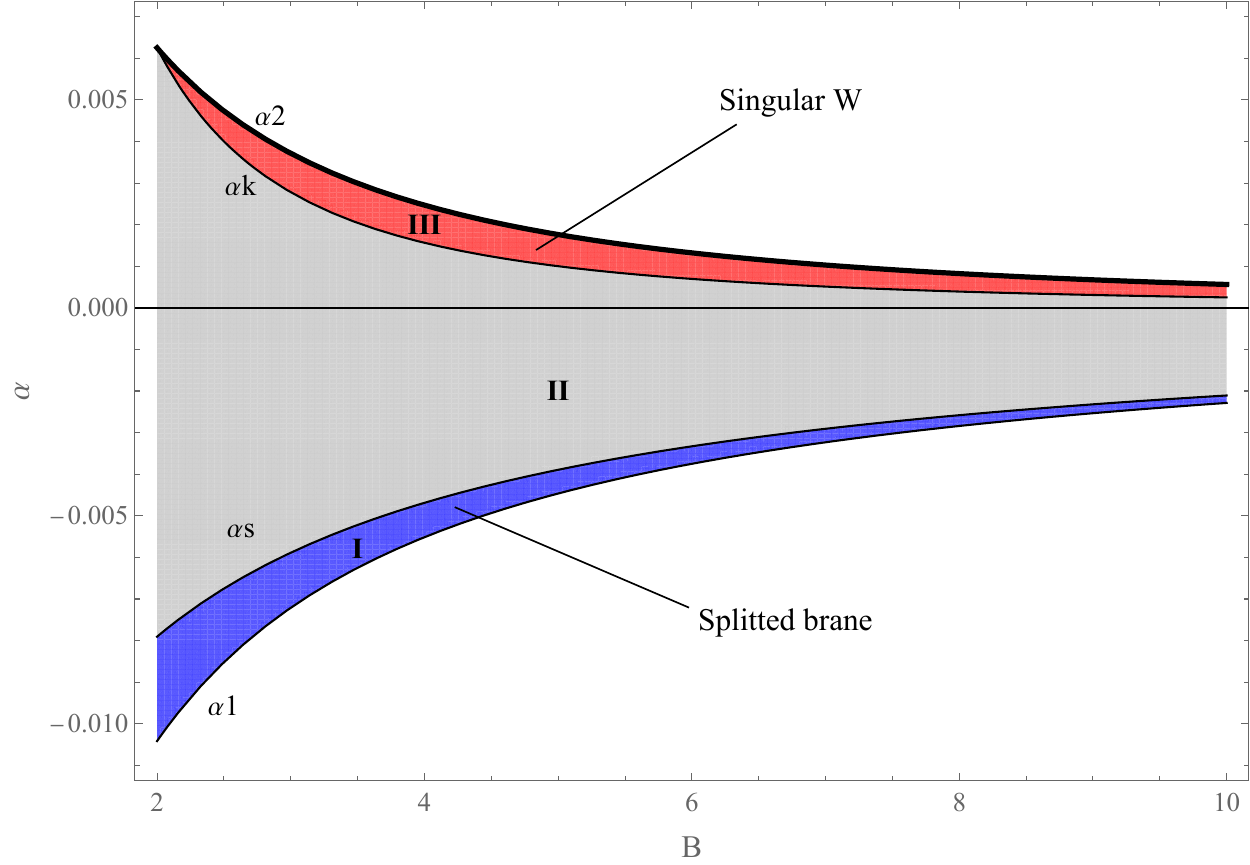}
\caption{\label{fig:pspace}
Parameter space in the \((\alpha,B)\) plane constrained by the scalar-field reality condition and the ghost-free condition \(f_R>0\). Region~I corresponds to the energy-density splitting regime, Region~II to the smooth ghost-free branch, and Region~III to the singular branch where \(f_R\) crosses zero.}
\end{figure}

The allowed regions in the \((\alpha,B)\) plane are summarized in figure~\ref{fig:pspace}. Region~I corresponds to the energy-density splitting regime \(\alpha_1\leq\alpha\leq\alpha_s\), where the brane develops an internal structure. Region~II denotes the smooth ghost-free branch with no singularity in \(f_R\). Region~III exists only for sufficiently thick branes, \(B>2\), and corresponds to the singular branch where \(f_R\) vanishes and the tensor effective potential diverges~\cite{Xu:2015}.

In summary, the two-scalar \(f(R)\) construction provides an analytical thick-brane background with tunable internal energy-density structure. The parameters \(B\) and \(\alpha\) control, respectively, the geometric localization of the warp factor and the higher-curvature modification of the background. However, once \(A(y)\) and \(f_R(y)\) are fixed, the tensor perturbation equation is determined by these background functions. The two scalar fields therefore act as an analytical matter source for generating the background structure, rather than introducing an additional direct coupling in the tensor sector. In the next section we derive the tensor perturbation equation and analyze how the smooth ghost-free branch and the singular \(f_R=0\) branch lead to different real-axis and complex-frequency spectral responses~\cite{Zhong:2011, Gu:2016nyo}.

\section{Formalism of tensor gravitational perturbations}
\label{sec:perturbation}

To investigate the tensor-sector stability and the massive Kaluza-Klein spectrum of the two-scalar \(f(R)\) thick brane, we consider linear tensor perturbations around the background metric~\eqref{eq:metric}~\cite{DeWolfe:1999cp, Csaki:2000fc}. The perturbed metric is written as
\begin{equation}
g_{MN} =
\begin{pmatrix}
e^{2A(y)} \left[ \eta_{\mu\nu} + h_{\mu\nu}(x^\rho,y) \right] & 0 \\
0 & 1
\end{pmatrix} \,.
\label{eq:perturbed_metric}
\end{equation}
We impose the transverse-traceless gauge conditions
\begin{equation}
\partial^\mu h_{\mu\nu}=0 \,,
\qquad
\eta^{\mu\nu}h_{\mu\nu}=0 \,.
\label{eq:TT_gauge}
\end{equation}
Under these conditions, the tensor modes decouple from the scalar and vector perturbations. Keeping the linear terms in the modified Einstein equations gives the tensor perturbation equation~\cite{Zhong:2011, Xu:2015, E:2026}
\begin{equation}
\left[
\partial_y^2
+\left(4A' + \frac{f_R'}{f_R}\right)\partial_y
+e^{-2A}\Box^{(4)}
\right]h_{\mu\nu}(x^\rho,y)=0 \,,
\label{eq:tensor_5D_y}
\end{equation}
where \(f_R=1+2\alpha R\), the prime denotes differentiation with respect to \(y\), and \(\Box^{(4)}=\eta^{\mu\nu}\partial_\mu\partial_\nu\).

We introduce the conformal coordinate \(z\) through
\begin{equation}
dz=e^{-A(y)}dy \,.
\label{eq:z_coordinate}
\end{equation}
This coordinate is convenient for writing the perturbation equation as a one-dimensional scattering problem. With the rescaling
\begin{equation}
h_{\mu\nu}(x^\rho,z)
=
e^{-\frac{3}{2}A(z)}f_R^{-1/2}(z)
\epsilon_{\mu\nu}(x^\rho)\Phi(z,t) \,,
\label{eq:ansatz_z}
\end{equation}
the first-derivative term is removed and the time-dependent perturbation equation becomes
\begin{equation}
\left(
\frac{\partial^2}{\partial t^2}
-\frac{\partial^2}{\partial z^2}
+W(z)
\right)\Phi(z,t)=0 \,.
\label{eq:td_wave_eq}
\end{equation}
After the harmonic decomposition \(\Phi(z,t)=e^{-i\omega t}\Psi(z)\), one obtains the Schrödinger-like equation
\begin{equation}
\left(
-\frac{d^2}{dz^2}+W(z)
\right)\Psi(z)=m^2\Psi(z) \,,
\label{eq:sch_eq}
\end{equation}
where \(m^2=\omega^2\) is the four-dimensional KK mass parameter. In terms of \(a(z)=e^{A(z)}\), the effective potential is~\cite{Xu:2015, E:2026}
\begin{equation}
W(z)
=
\frac{3}{4}\frac{(\partial_z a)^2}{a^2}
+\frac{3}{2}\frac{\partial_z^2 a}{a}
+\frac{3}{2}\frac{(\partial_z a)(\partial_z f_R)}{a f_R}
-\frac{1}{4}\frac{(\partial_z f_R)^2}{f_R^2}
+\frac{1}{2}\frac{\partial_z^2 f_R}{f_R} \,.
\label{eq:potential_Wz}
\end{equation}

The tensor stability in the smooth ghost-free branch follows from the
supersymmetric factorization of the Schr\"odinger operator. Defining
\begin{equation}
\mathcal{W}(z)
=
\frac{3}{2}\frac{\partial_z a}{a}
+\frac{1}{2}\frac{\partial_z f_R}{f_R} \,,
\label{eq:superpotential}
\end{equation}
we can write
\begin{equation}
\mathcal{H}
=
-\frac{d^2}{dz^2}+W(z)
=
\mathcal{K}\mathcal{K}^\dagger ,
\qquad
\mathcal{K}=\partial_z+\mathcal{W}(z),
\qquad
\mathcal{K}^\dagger=-\partial_z+\mathcal{W}(z),
\label{eq:factorization}
\end{equation}
with
\begin{equation}
W(z)=\mathcal{W}^2(z)+\partial_z\mathcal{W}(z).
\label{eq:potential_factorized}
\end{equation}
Therefore, the Schr\"odinger-like equation~\eqref{eq:sch_eq} can be written as
\begin{equation}
\mathcal{K}\mathcal{K}^\dagger \Psi_m(z)=m^2\Psi_m(z).
\label{eq:kk_factorized}
\end{equation}
This form makes the non-negativity of the tensor spectrum explicit: as long as
\(f_R>0\) and the factorization is well defined, one has \(m^2\geq0\), and
tachyonic tensor modes are excluded~\cite{Kobayashi:2001jd, Zhong:2011}.

The supersymmetric partner potential is
\begin{equation}
W_s(z)
=
\mathcal{W}^2(z)-\frac{d\mathcal{W}(z)}{dz} \,.
\label{eq:partner_potential}
\end{equation}
This partner problem will be useful in the time-domain analysis, because the
partner potential removes the localized zero-mode contribution while preserving
the positive part of the supersymmetric spectrum under the usual assumptions.

The massless graviton zero mode follows from
\[
\mathcal{K}^\dagger \Psi_0=0
\]
and is given by
\begin{equation}
\Psi_0(z)
=
N_0\, a^{3/2}(z) f_R^{1/2}(z) \,,
\label{eq:zero_mode}
\end{equation}
where \(N_0\) is a normalization constant. When \(f_R>0\) globally, this zero
mode is normalizable for the backgrounds considered here. Its localization is
responsible for the recovery of four-dimensional Newtonian gravity on the
brane~\cite{Randall:1999vf, Csaki:2000fc}.

For generic \(B\), the coordinate relation \(z(y)\) cannot be inverted
analytically. Therefore, for plotting and numerical implementation, it is useful
to rewrite the potential as a function of the physical coordinate \(y\). Using
\(\partial_z=e^{A(y)}\partial_y\), one obtains
\begin{equation}
W(z(y))
=
e^{2A(y)}
\left[
\frac{15}{4}(\partial_y A)^2
+\frac{3}{2}\partial_y^2 A
+2\frac{(\partial_y A)(\partial_y f_R)}{f_R}
-\frac{1}{4}\frac{(\partial_y f_R)^2}{f_R^2}
+\frac{1}{2}\frac{\partial_y^2 f_R}{f_R}
\right] .
\label{eq:potential_Wy}
\end{equation}

\begin{figure}[tbp]
\centering
\includegraphics[width=0.95\textwidth]{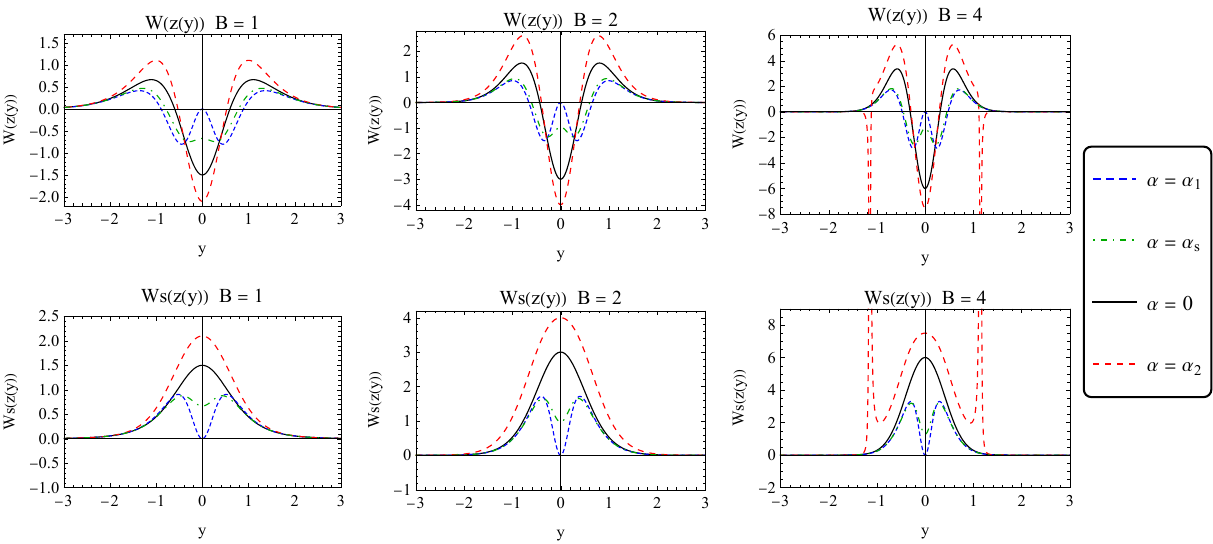}
\caption{\label{fig:potentials_2d}
Effective potentials \(W(z(y))\) and \(W_s(z(y))\) for \(B=1,2,4\) and
representative values of \(\alpha\). The potential \(W\) is evaluated from
eq.~\eqref{eq:potential_Wy}, while the partner potential \(W_s\) is obtained
from eq.~\eqref{eq:partner_potential}. The potentials are plotted as functions
of the physical coordinate \(y\). In the smooth ghost-free region they are
regular and asymptotically vanish, while in the singular branch \(f_R=0\)
produces divergent structures.}
\end{figure}

The profiles of the effective potential \(W(z(y))\) and its supersymmetric
partner \(W_s(z(y))\) are shown in figure~\ref{fig:potentials_2d}.In the ghost-free regions, the potentials are smooth and approach zero at spatial infinity. Hence the massive tensor spectrum is continuous and gapless, as in standard infinite-extra-dimension thick-brane models~\cite{Randall:1999vf, Gremm:1999pj}. The parameters \(B\) and \(\alpha\) control the width and the height of the potential barriers. In particular, the energy-density splitting near \(\alpha_s\) can modify the central structure of the potential.

A qualitatively different behavior appears in the singular branch. For \(B>2\) and \(\alpha_k\leq\alpha\leq\alpha_2\), the effective coupling \(f_R\) crosses zero. Since eq.~\eqref{eq:potential_Wy} contains inverse powers of \(f_R\), the effective potential develops singular structures. These singularities should not be confused with ordinary smooth trapping wells: they signal the breakdown of the ghost-free condition and must be physically distinguished from the smooth branch. In the following real-axis analysis, we will see that sharp quasi-localization peaks appear only in this singular branch.

We now examine whether the massive tensor modes form quasi-localized structures on the real mass axis. For a massive mode, the wave function is not normalizable on the infinite extra dimension and behaves asymptotically as a scattering state. A commonly used real-axis diagnostic is the relative probability method~\cite{Liu:2009ve, Almeida:2009jc, Tan:2023cra}. It measures the fraction of the wave-function amplitude concentrated near the brane:
\begin{equation}
P(m^2)
=
\frac{
\int_{-z_b}^{z_b}|\Psi_m(z)|^2 dz
}{
\int_{-z_{\max}}^{z_{\max}}|\Psi_m(z)|^2 dz
} \,,
\label{eq:relative_prob}
\end{equation}
where \(2z_b\) characterizes the brane region and \(z_{\max}\) is the numerical cutoff.

The relative-probability method should be understood as a diagnostic for narrow, long-lived quasi-localized modes on the real mass axis. A sharp Lorentzian-like peak in \(P(m^2)\) indicates that the corresponding KK mode spends a comparatively long time near the brane before leaking into the bulk. The peak width is then related to the lifetime of the quasi-localized state. However, this method does not locate broad poles far from the real axis. Therefore, the absence of sharp peaks in \(P(m^2)\) should not be interpreted as the absence of all spectral structures; it only means that no long-lived tensor mode is resolved on the real mass axis by this criterion.

\begin{figure}[tbp]
\centering
\includegraphics[width=1\textwidth]{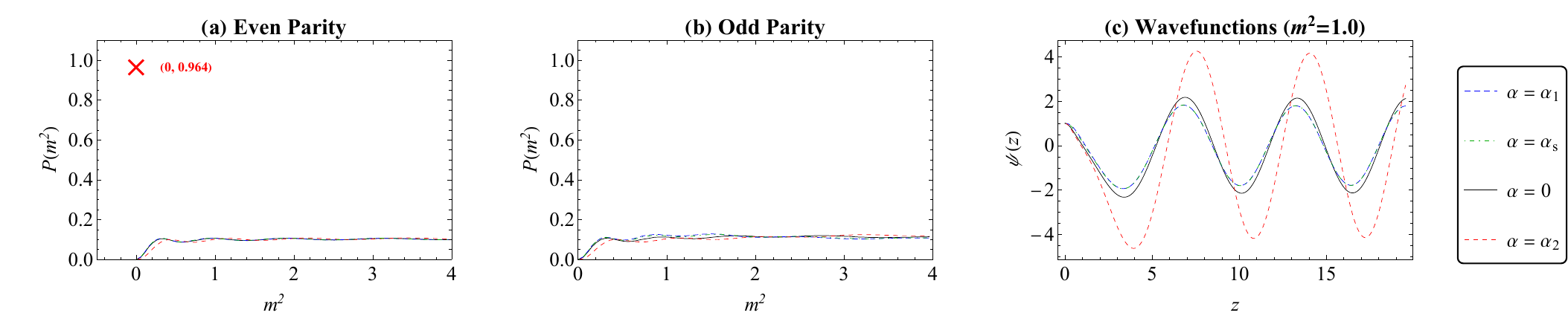}
\caption{\label{fig:resonance_stable}
Relative probability \(P(m^2)\) defined in eq.~\eqref{eq:relative_prob} for even parity (left) and odd parity (center) modes with \(B=2\). The right panel shows representative even-parity wave functions governed by eq.~\eqref{eq:sch_eq} at \(m^2=1\). No narrow real-axis peaks are resolved in the smooth ghost-free branch.}
\end{figure}

We first consider the representative ghost-free case \(B=2\). The relative-probability spectra are shown in figure~\ref{fig:resonance_stable}. For all tested values of \(\alpha\), including the splitting point \(\alpha_s\), the curves remain smooth and low in both parity sectors. No narrow peaks are resolved. The representative wave functions at \(m^2=1\) are extended over the extra dimension and do not show a pronounced enhancement near the brane. This indicates that the smooth ghost-free potentials are too transparent to support long-lived tensor modes detectable by the relative-probability criterion. In particular, the energy-density splitting by itself does not generate a narrow tensor resonance in the ghost-free region.

To check whether the absence of sharp peaks is merely a byproduct of the relative-probability diagnostic, we also compute the real-axis scattering phase shifts. For each real mass \(m\), eq.~\eqref{eq:sch_eq} is solved in the even and odd parity channels with
\begin{equation}
\psi_e(0)=1,\qquad \psi'_e(0)=0,
\qquad
\psi_o(0)=0,\qquad \psi'_o(0)=1 .
\label{eq:phase_initial_conditions}
\end{equation}
Since \(W(z)\to0\) at large \(|z|\), the asymptotic behavior can be fitted as
\begin{equation}
\psi_e(z)\simeq A_e\cos(mz+\delta_e),
\qquad
\psi_o(z)\simeq A_o\sin(mz+\delta_o),
\label{eq:phase_shift_asymptotic}
\end{equation}
where \(\delta_e\) and \(\delta_o\) are the even and odd phase shifts. For a symmetric one-dimensional potential, the transmission coefficient is
\begin{equation}
\mathcal{T}(m)
=
\cos^2\!\left[\delta_e(m)-\delta_o(m)\right].
\label{eq:transmission_coeff}
\end{equation}

\begin{figure}[tbp]
\centering
\includegraphics[width=0.95\textwidth]{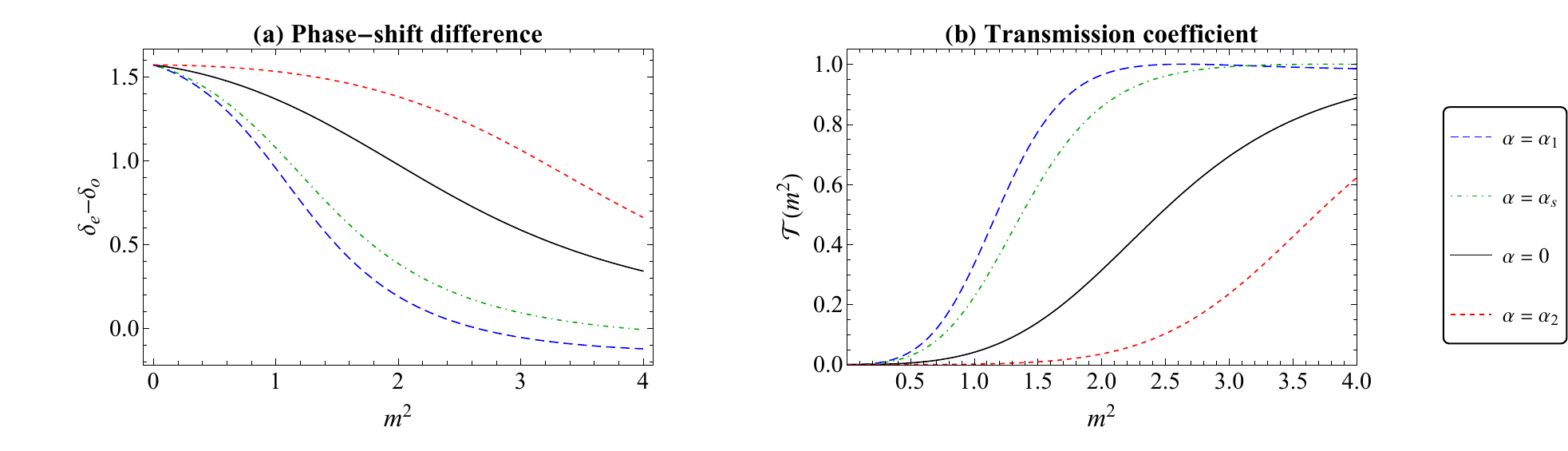}
\caption{\label{fig:transmission}
Real-axis scattering diagnostic for the representative case \(B=2\). Left: the phase-shift difference \(\delta_e-\delta_o\) extracted from the asymptotic even and odd parity wave functions. Right: the corresponding transmission coefficient \(\mathcal{T}(m)=\cos^2(\delta_e-\delta_o)\), plotted as a function of \(m^2\). The transmission coefficient increases smoothly and shows no narrow resonant peaks, consistently with the relative-probability spectra in figure~\ref{fig:resonance_stable}. The curve at \(\alpha=\alpha_2\) is included as the limiting boundary value for \(B=2\).}
\end{figure}

The phase-shift difference and transmission coefficient for \(B=2\) are shown in figure~\ref{fig:transmission}. The transmission coefficient grows smoothly with \(m^2\), as expected for ordinary barrier scattering. No narrow transmission peaks are found for any of the tested values of \(\alpha\). This result independently supports the conclusion from the relative-probability analysis: in the smooth ghost-free region, the real-axis scattering response does not exhibit long-lived tensor resonances. The corresponding massive KK excitations should therefore be characterized by their complex QNM frequencies rather than by narrow real-axis peaks.

\begin{figure}[tbp]
\centering
\includegraphics[width=0.95\textwidth]{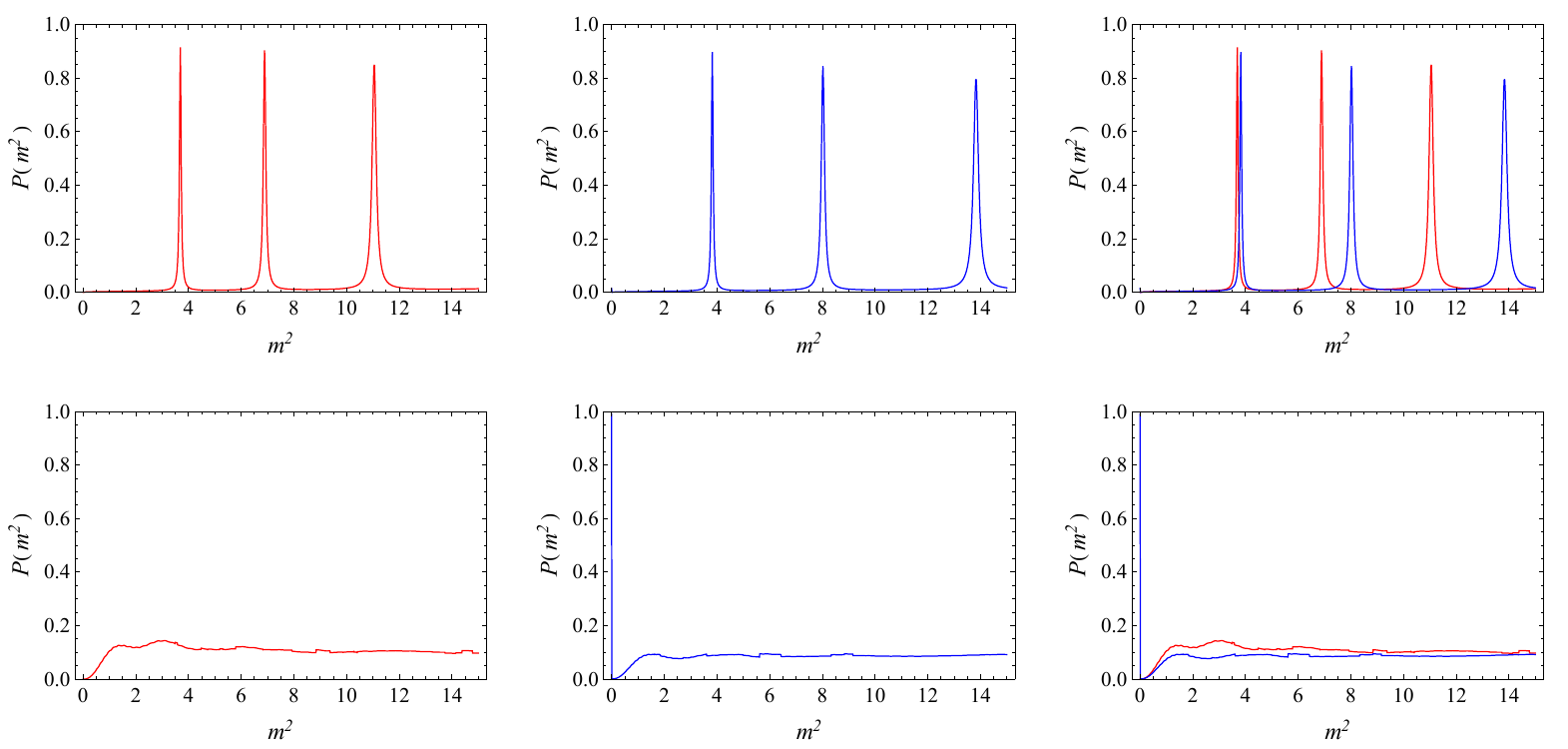}
\caption{\label{fig:ghost_spectra}
Relative-probability spectra for \(B=4\). Top row: spectra at \(\alpha=\alpha_2\), where \(f_R\) crosses zero and the effective potential develops singular structures. Sharp quasi-localization peaks appear in this singular branch. Bottom row: spectra at the splitting point \(\alpha=\alpha_s\), where the brane has an internal structure but no narrow real-axis peaks are resolved. Red and blue curves denote odd and even parity modes, respectively.}
\end{figure}

The situation changes when the parameters enter the singular branch. For \(B>2\), the upper ghost-free bound is \(\alpha_k=1/(40B^2)\). When \(\alpha_k\leq\alpha\leq\alpha_2\), the effective coupling \(f_R\) crosses zero and the inverse powers of \(f_R\) in \(W(z)\) generate singular structures. Figure~\ref{fig:ghost_spectra} illustrates the contrast for \(B=4\). At the singular boundary \(\alpha=\alpha_2\), sharp peaks appear in the relative-probability spectrum. By contrast, at the Bloch-brane splitting point \(\alpha=\alpha_s\), the spectra remain low and smooth, even though the energy density has an internal double-peak structure. This comparison shows that the sharp real-axis peaks are associated with the singular \(f_R=0\) boundary rather than with the matter-sector splitting alone.

\begin{figure}[tbp]
\centering
\includegraphics[width=0.95\textwidth]{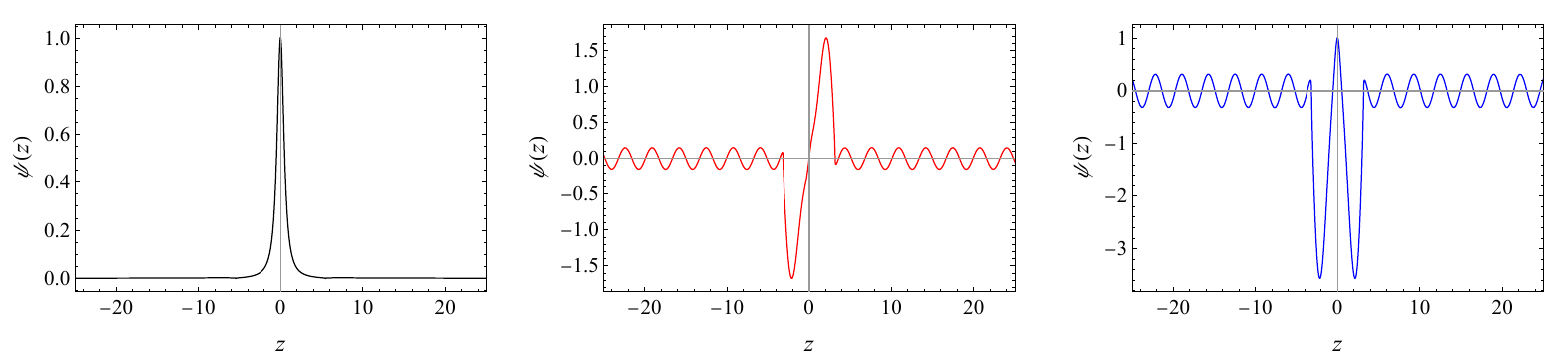}
\caption{\label{fig:ghost_wavefunctions}
Representative wave functions in the singular branch with \(B=4\) and \(\alpha=\alpha_2\). The panels show the formal zero-mode profile (left) and the first odd- and even-parity modes associated with the sharp relative-probability peaks (center and right). These modes are enhanced near the brane region, reflecting singular-boundary quasi-localization rather than trapping by a smooth ghost-free potential.}
\end{figure}

The corresponding wave functions in the singular branch are displayed in figure~\ref{fig:ghost_wavefunctions}. The modes associated with the sharp peaks are strongly enhanced near the brane region, in contrast to the extended scattering states found in the smooth ghost-free branch. These profiles provide a useful diagnostic of the singular-boundary quasi-localization effect. Since the background violates the ghost-free condition in this branch, however, these sharp structures should not be interpreted as ghost-free long-lived resonances of the smooth brane model.

The real-axis analysis therefore leads to a clear separation. In the smooth ghost-free region, both the relative-probability and transmission diagnostics show no narrow tensor resonances. The massive KK modes are broad and rapidly leaking states. In the singular branch, sharp real-axis peaks can appear, but they are tied to the zero of \(f_R\) and the resulting singular potential. The absence of narrow real-axis peaks in the ghost-free region does not imply a trivial spectrum; rather, it indicates that the relevant poles are broad and should be analyzed in the complex frequency plane. This motivates the QNM analysis in the next section.

\section{Quasinormal modes and time-domain dynamics}
\label{sec:qnm}

The real-axis analysis in section~\ref{sec:perturbation} shows that the smooth
ghost-free branch does not support narrow tensor resonances detectable by the
relative-probability or transmission diagnostics. This does not mean that the
massive KK spectrum is trivial. Rather, it indicates that the relevant spectral
structures are broad and dissipative, corresponding to poles away from the real
mass axis. From the scattering-theory viewpoint, resonances and QNMs are related
descriptions of the pole structure of an open wave system~\cite{Berti:2009kk,
Konoplya:2011qq}. Narrow real-axis peaks correspond to long-lived modes with
small damping rates, while short-lived modes are better characterized directly
by their complex frequencies.

We therefore analyze the ghost-free branch in the complex frequency plane. The
goal of this section is to extract the QNFs and to show that the massive tensor
excitations in the smooth branch are strongly damped, short-lived modes. We use
frequency-domain calculations where applicable and complement them with
time-domain evolutions. The latter are particularly useful for backgrounds where
the coordinate transformation \(z(y)\) cannot be inverted analytically. In the
time-domain calculation we also use the supersymmetric partner potential as a
numerical device to reduce the contamination from the non-dissipative graviton
zero mode.

\subsection{Asymptotic Iteration Method and quasinormal mode spectra}
\label{subsec:aim}

We start from the harmonic decomposition
\[
\Phi(z,t)=e^{-i\omega t}\Psi(z)
\]
in the time-dependent wave equation~\eqref{eq:td_wave_eq}. The resulting
Schr\"{o}dinger-like equation is
\begin{equation}
\left[
-\frac{d^2}{dz^2}+W(z)
\right]\Psi(z)=\omega^2\Psi(z) \,.
\label{eq:qnm_sch}
\end{equation}
The QNM frequency is complex,
\begin{equation}
\omega=\omega_R+i\omega_I \,.
\end{equation}
The real part \(\omega_R\) gives the oscillation scale of the ringdown, while
\(\omega_I\) determines the damping rate. With the convention \(e^{-i\omega t}\),
a stable damped mode has \(\omega_I<0\). For an open system, the wave function
satisfies purely outgoing boundary conditions at spatial infinity~\cite{Kokkotas:1999bd,
Nollert:1999ji},
\begin{equation}
\Psi(z)\sim e^{\pm i\omega z}\,,\qquad z\rightarrow \pm\infty .
\label{eq:qnm_bc}
\end{equation}

For cases where the potential is analytically or semi-analytically tractable in
the conformal coordinate, we compute QNFs using the Asymptotic Iteration Method
(AIM). We map the infinite domain \(z\in(-\infty,\infty)\) to a finite interval
\(u\in(-1,1)\) by
\[
u=\frac{\sqrt{4k^2z^2+1}-1}{2kz}.
\]
After factoring out the asymptotic outgoing behavior in
eq.~\eqref{eq:qnm_bc}, the wave function is written as
\[
\Psi(u)=\mathcal{A}(u)\chi(u),
\]
where \(\mathcal{A}(u)\) contains the boundary factors. The perturbation equation
is then recast into the AIM form~\cite{Ciftci:2003As, AIM_2011}
\begin{equation}
\chi''(u)=\lambda_0(u,\omega)\chi'(u)+s_0(u,\omega)\chi(u).
\label{eq:aim_entry}
\end{equation}
Successive differentiations lead to recursion coefficients \(\lambda_n\) and
\(s_n\), and the QNFs are obtained from the quantization condition
\begin{equation}
s_n(u,\omega)\lambda_{n-1}(u,\omega)
-
s_{n-1}(u,\omega)\lambda_n(u,\omega)=0 .
\label{eq:aim_quantization}
\end{equation}

\begin{figure}[tbp]
\centering
\includegraphics[width=0.6\textwidth]{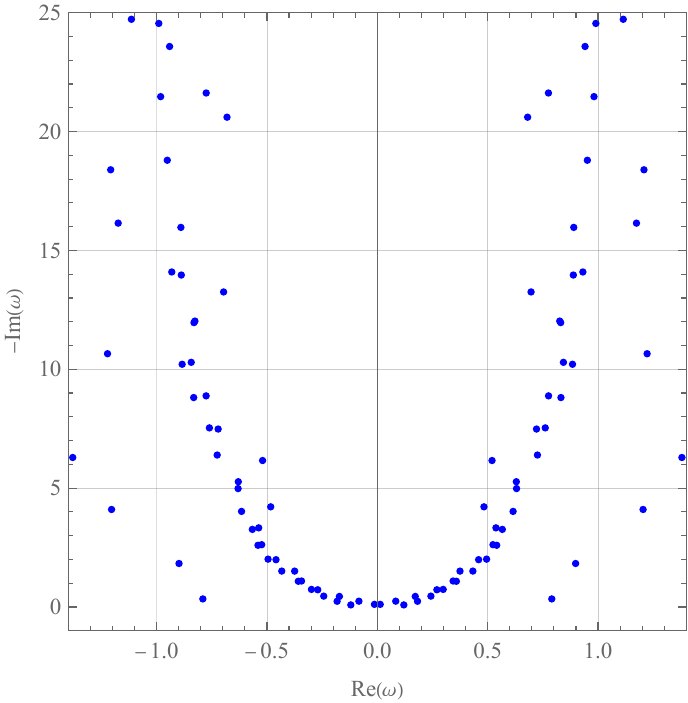}
\caption{\label{fig:aim_scatter}
Distribution of the quasinormal frequencies in the complex plane obtained via
the AIM with iteration number \(n=340\), for \(B=1\) and \(\alpha=\alpha_1\).
All displayed modes have \(\omega_I<0\), corresponding to damped ringdown
signals.}
\end{figure}

Figure~\ref{fig:aim_scatter} shows a representative AIM spectrum for
\(B=1\) and \(\alpha=\alpha_1\). The extracted modes lie in the lower half of
the complex-frequency plane, \(\omega_I<0\), indicating damped time evolution.
This is consistent with the tensor-sector stability implied by the
supersymmetric factorization of the Schr\"{o}dinger operator in the ghost-free
region.

\subsection{Time-domain evolution and zero-mode elimination}
\label{subsec:zero_mode}

To complement the frequency-domain calculation and to visualize the dissipative
dynamics, we solve the time-dependent equation~\cite{Press1971, Cunningham1978}
\begin{equation}
\left(
\frac{\partial^2}{\partial t^2}
-\frac{\partial^2}{\partial z^2}
+W(z)
\right)\Phi(z,t)=0 .
\label{eq:td_eq}
\end{equation}
The initial perturbation is chosen as a Gaussian wave packet localized away from
the brane and propagating toward the potential region. The subsequent signal
contains the scattering response of the brane potential and, in suitable time
windows, the QNM ringdown.

A technical issue in the original potential \(W(z)\) is the presence of the
localized massless graviton zero mode. This mode is responsible for recovering
four-dimensional Newtonian gravity on the brane~\cite{Csaki:2000fc, Liu:2017gcn},
but it is non-dissipative. In time-domain simulations with even-parity initial
data, the zero mode can leave a persistent late-time component near the brane.
This component is not a QNM of the massive sector and can obscure the extraction
of the dissipative ringdown at late times.

\begin{figure}[tbp]
\centering
\includegraphics[width=0.92\textwidth]{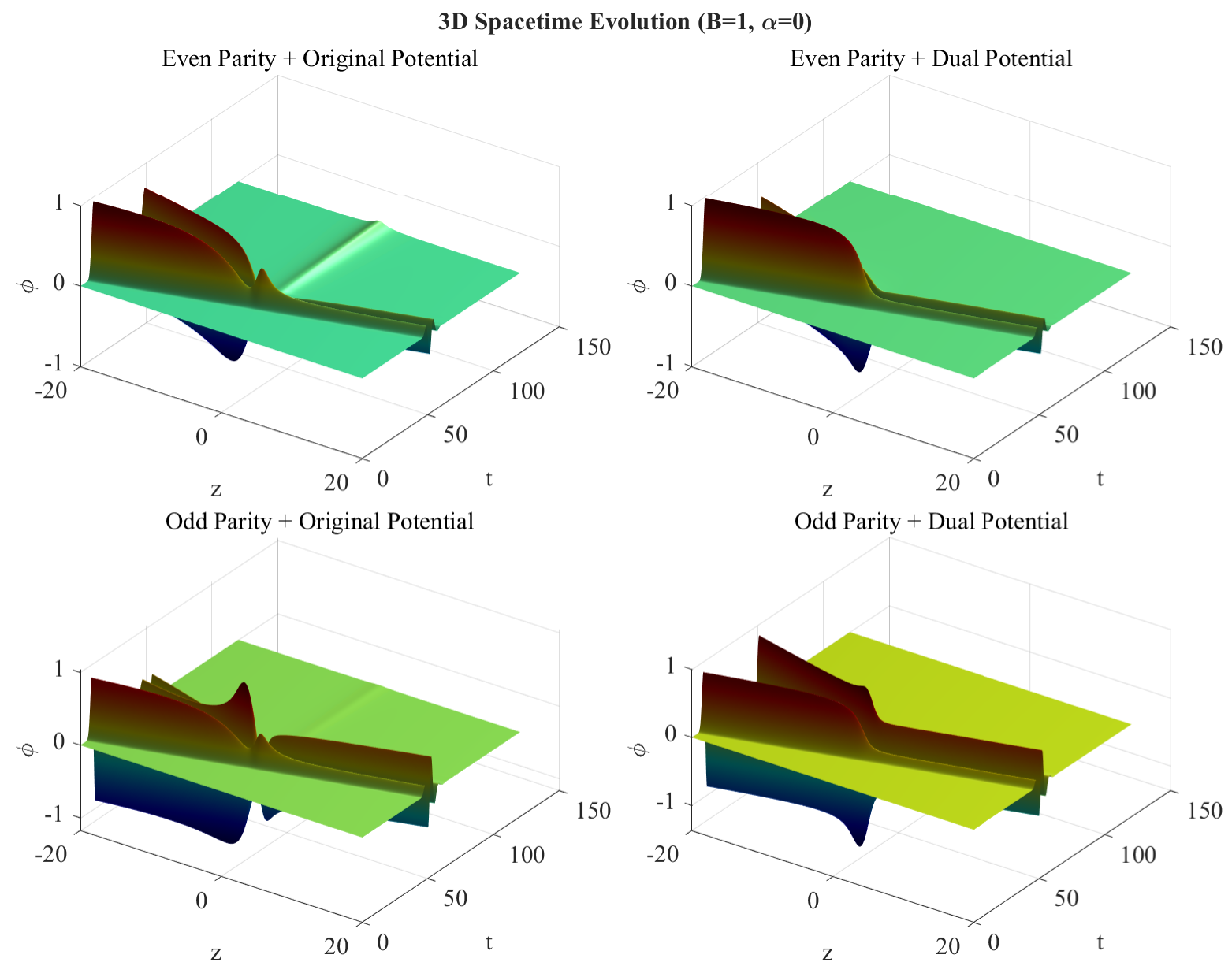}
\caption{\label{fig:evolution_3d}
Three-dimensional spatiotemporal evolution of the wave-packet amplitude
\(|\Phi(z,t)|\) governed by eq.~\eqref{eq:td_eq}, for different choices of the
effective potential and the initial parity. The initial Gaussian packet is
localized around \(kz=-30\). The plotted window \(kz\in[-20,20]\) highlights
the scattering process near the brane region and the subsequent propagation of
the wave packet into the bulk.}
\end{figure}

\begin{figure}[tbp]
\centering
\includegraphics[width=0.9\textwidth]{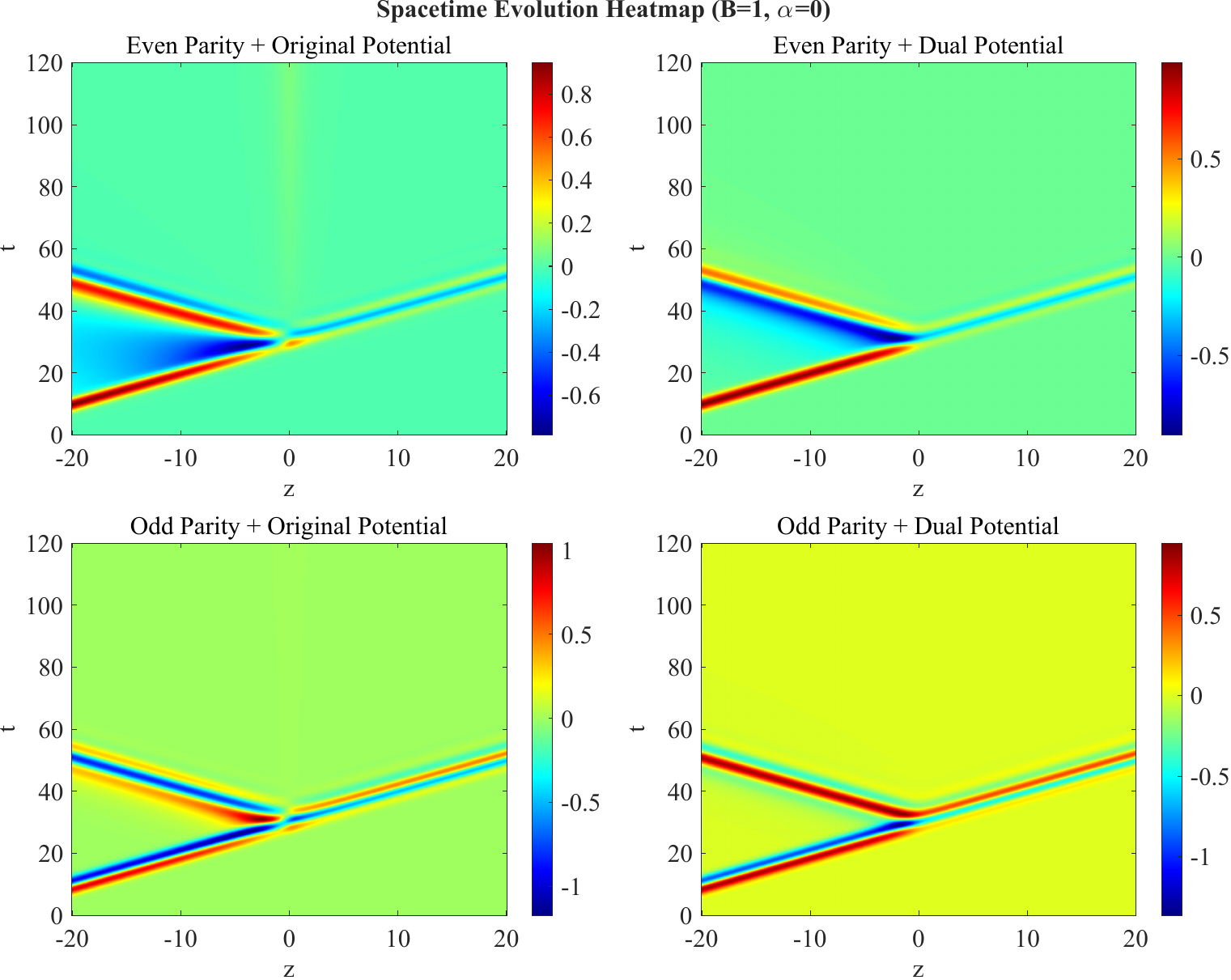}
\caption{\label{fig:evolution_heatmap}
Heatmap representations corresponding to the evolutions in
figure~\ref{fig:evolution_3d}. With the original potential \(W(z)\) and
even-parity initial data, a persistent zero-mode component remains near
\(kz=0\). In the partner-potential evolutions, this localized non-dissipative
component is removed, providing a cleaner time window for extracting the
dissipative ringdown signal.}
\end{figure}

To reduce this contamination, we compare four evolution strategies: even- and
odd-parity initial data evolved under the original potential \(W(z)\), and the
same two types of initial data evolved under the supersymmetric partner
potential \(W_s(z)\). The comparison is shown in
figures~\ref{fig:evolution_3d} and~\ref{fig:evolution_heatmap}. Under the
original potential with even-parity initial data, a persistent localized band is
visible near the brane position. Odd-parity initial data suppresses the direct
excitation of the even zero mode, although small numerical contamination may
still appear at very late times.

The partner potential \(W_s(z)\) does not contain the same localized zero-mode
bound state. Evolving the wave packet under \(W_s(z)\) therefore provides a
cleaner numerical setup for isolating the dissipative part of the response. We
use this partner-potential evolution mainly as a zero-mode filtering tool. The
consistency of the extracted QNFs between the original and partner potentials is
checked explicitly in table~\ref{tab:qnm_comprehensive}.

\subsection{Signal extraction and methodological cross-validation}
\label{subsec:cross_validation}

\begin{figure}[tbp]
\centering
\includegraphics[width=0.95\textwidth]{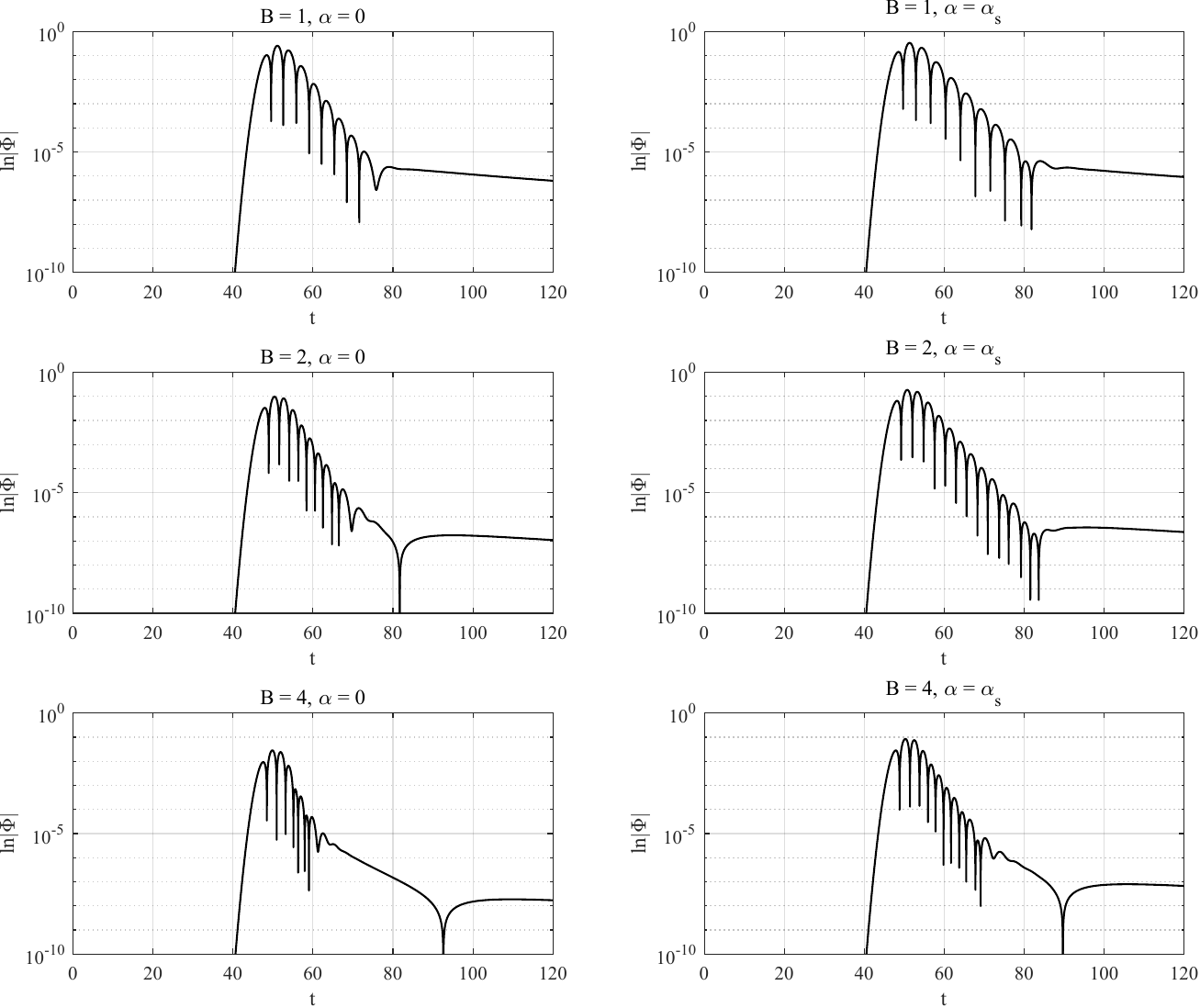}
\caption{\label{fig:td_modulation}
Time-domain ringdown amplitudes \(|\Phi(t,z=20)|\), plotted on a logarithmic
vertical scale, for \(B=1,2,4\). The left column corresponds to \(\alpha=0\),
and the right column to \(\alpha=\alpha_s\). Increasing \(B\) raises the
typical oscillation frequency, while changing \(\alpha\) modifies the damping
rate through the corresponding change of the effective potential.}
\end{figure}

We now use the time-domain waveforms to extract the fundamental QNFs and to
study how the parameters \(B\) and \(\alpha\) affect the ringdown. The signals
shown in figure~\ref{fig:td_modulation} are measured at \(z=20\). The vertical
axis is plotted on a logarithmic scale. A larger value of \(B\) generally makes
the effective potential narrower and steeper, which increases the oscillation
scale of the ringdown. Changing \(\alpha\) modifies the height and shape of the
potential barriers, thereby changing both the oscillation frequency and the
damping rate.

\begin{figure}[tbp]
\centering
\includegraphics[width=1\textwidth]{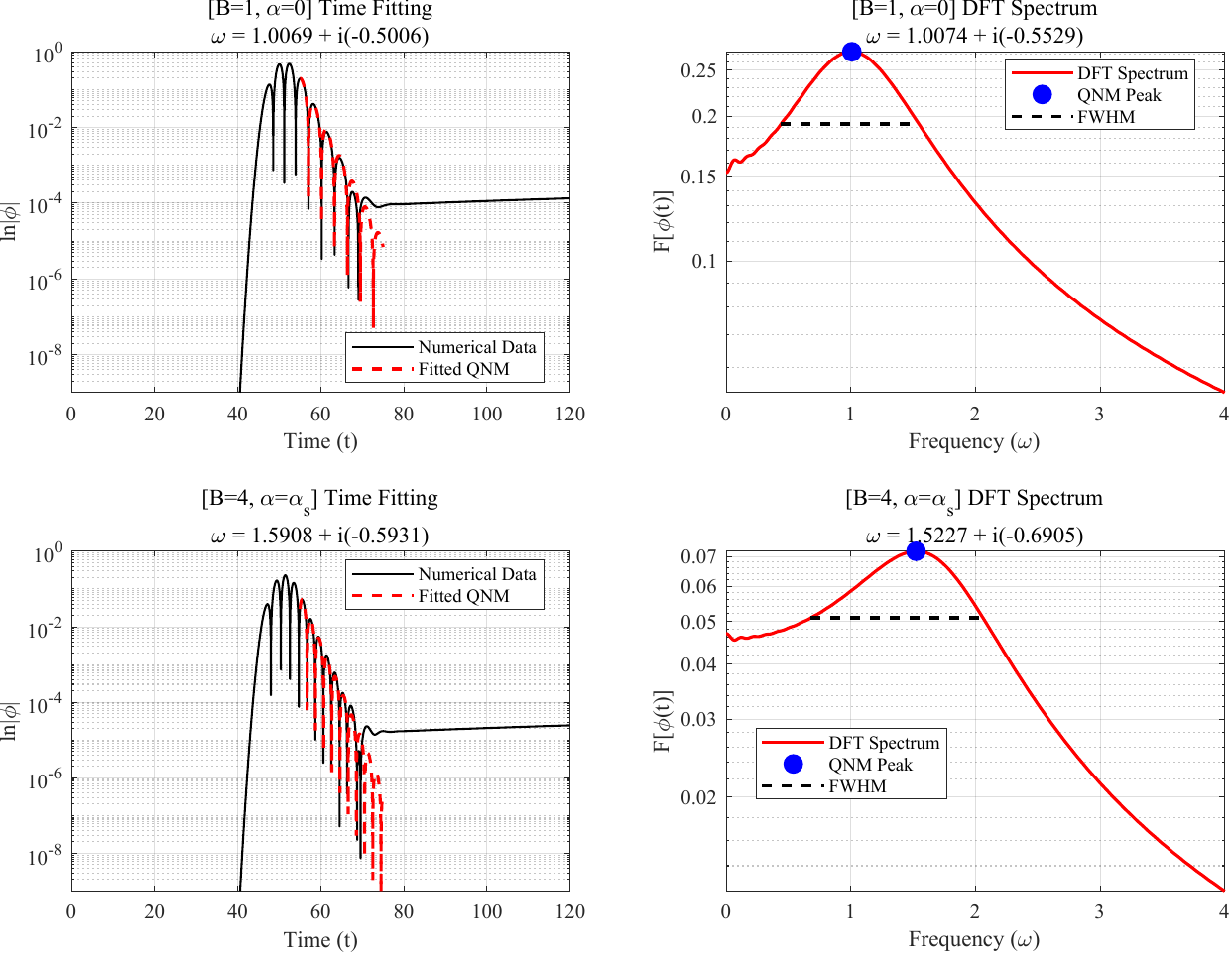}
\caption{\label{fig:extraction_fft}
Demonstration of QNF extraction from time-domain data. Left: exponential
fitting of the ringdown envelope on a logarithmic vertical scale to determine
the damping rate \(\omega_I\). Right: DFT spectrum of the waveform, where the
dominant peak estimates the oscillation frequency \(\omega_R\).}
\end{figure}

The complex QNF can be extracted from the time-domain data by fitting the
ringdown signal to a damped sinusoidal form,
\begin{equation}
\Phi(t,z_{\rm obs})
\simeq
A\,e^{\omega_I t}\cos(\omega_R t+\varphi),
\qquad
\omega_I<0 .
\label{eq:td_fit_form}
\end{equation}
We also use the discrete Fourier transform (DFT) of the waveform as an
independent check of the dominant oscillation frequency~\cite{Berti:2009kk,
DIM_2013}. Figure~\ref{fig:extraction_fft} illustrates the two extraction
procedures. The exponential envelope gives the damping rate, while the dominant
peak in the DFT spectrum gives the oscillation frequency.

Table~\ref{tab:qnm_comprehensive} summarizes the fundamental QNFs obtained from
frequency-domain calculations and from the four time-domain evolution
strategies. The time-domain results obtained with different initial parities and
different evolution potentials agree at the level needed for the present
spectral classification. This agreement indicates that the extracted QNFs are
not artifacts of a specific initial wave packet or of the zero-mode filtering
procedure.

The time-domain approach is also useful beyond the cases where the AIM is
straightforward to implement. For \(B\neq1\), the lack of a closed-form
coordinate inversion \(z(y)\) makes a direct AIM implementation less convenient
without additional numerical treatment~\cite{Xu:2015, Tan:2023cra}. The
time-domain evolution works directly with numerical grids and therefore provides
a practical complementary method for general \(B\).

The configuration \(B=4,\alpha=\alpha_2\) is excluded from the QNM table because
it belongs to the singular branch rather than to the smooth ghost-free scattering
problem. In this case \(f_R\) crosses zero and the effective potential develops
singular structures. The standard QNM extraction used for smooth open potentials
is therefore not well defined in the same sense. The sharp real-axis peaks found
in section~\ref{sec:perturbation} for this configuration should be interpreted
as singular-boundary quasi-localization signals, not as ordinary QNMs of a
ghost-free smooth background.

\begin{table}[tbp]
\centering
\caption{\label{tab:qnm_comprehensive}
Cross-validation of the fundamental quasinormal frequencies
\(\omega=\omega_R+i\omega_I\). Results are compared between the
frequency-domain calculations and four time-domain strategies. Here \(P\) and
\(DP\) denote the original potential \(W(z)\) and the partner potential
\(W_s(z)\), respectively. The symbol ``--'' denotes cases where the method is
not applied, either because of analytical limitations or because the
configuration lies in the singular branch.}
\vspace{0.2cm}
\resizebox{\textwidth}{!}{ 
\begin{tabular}{ccccccc}
\hline\hline
$B$ & $\alpha$ & Frequency-Domain & P \& Odd-Wave & DP \& Odd-Wave & P \& Even-Wave & DP \& Even-Wave \\
\hline
\multirow{4}{*}{1} 
& $0$        & $0.9970 - 0.5263i^a$ & $0.9957 - 0.5052i$ & $0.9941 - 0.5046i$ & $0.9896 - 0.5108i$ & $0.9761 - 0.5217i$ \\
& $\alpha_1$ & $0.7903 - 0.3386i^a$ & $0.7970 - 0.3344i$ & $0.7967 - 0.3336i$ & $0.7938 - 0.3341i$ & $0.7874 - 0.3334i$ \\
& $\alpha_2$ & $1.3005 - 0.5460i^a$ & $1.2941 - 0.5401i$ & $1.2943 - 0.5398i$ & $1.2902 - 0.5331i$ & $1.2962 - 0.5423i$ \\
& $\alpha_s$ & $0.8428 - 0.3998i^a$ & $0.8509 - 0.3903i$ & $0.8502 - 0.3893i$ & $0.8454 - 0.3912i$ & $0.8361 - 0.3922i$ \\
\hline
\multirow{4}{*}{2} 
& $0$        & $1.5108 - 0.6309i^b$ & $1.4905 - 0.6480i$ & $1.4912 - 0.6503i$ & $1.5131 - 0.6502i$ & $1.5322 - 0.6535i$ \\
& $\alpha_1$ & --                   & $1.1079 - 0.3990i$ & $1.1069 - 0.3994i$ & $1.1081 - 0.4067i$ & $1.1081 - 0.4123i$ \\
& $\alpha_2$ & $1.8954 - 0.5121i^b$ & $1.8724 - 0.5269i$ & $1.8729 - 0.5285i$ & $1.8915 - 0.5270i$ & $1.8920 - 0.5270i$ \\
& $\alpha_s$ & $1.1756 - 0.4609i^b$ & $1.1650 - 0.4547i$ & $1.1639 - 0.4556i$ & $1.1677 - 0.4651i$ & $1.1694 - 0.4728i$ \\
\hline
\multirow{4}{*}{4} 
& $0$        & $2.2098 - 0.7991i^b$ & $2.2955 - 0.7914i$ & $2.2998 - 0.7823i$ & $2.1843 - 0.7011i$ & $2.0649 - 0.6578i$ \\
& $\alpha_1$ & --                   & $1.5563 - 0.5361i$ & $1.5575 - 0.5376i$ & $1.5745 - 0.5301i$ & $1.5855 - 0.5269i$ \\
& $\alpha_2$ & --                   & --                 & --                 & --                 & --                 \\
& $\alpha_s$ & --                   & $1.6163 - 0.5912i$ & $1.6182 - 0.5929i$ & $1.6396 - 0.5794i$ & $1.6542 - 0.5728i$ \\
\hline\hline
\multicolumn{7}{l}{\footnotesize $^a$ Calculated via the AIM based on eq.~\eqref{eq:sch_eq} in this work.} \\
\multicolumn{7}{l}{\footnotesize $^b$ Benchmarked against the direct integration method in ref.~\cite{E:2026}.} \\
\end{tabular}
}
\end{table}

It is useful to quantify the damping strength of these modes by the quality
factor
\begin{equation}
Q=\frac{\omega_R}{2|\omega_I|}.
\label{eq:q_factor}
\end{equation}
For the fundamental modes listed in table~\ref{tab:qnm_comprehensive}, we find
\(Q\simeq0.9-1.9\) in the ghost-free parameter region. Equivalently, after one
oscillation period \(T_{\rm osc}=2\pi/\omega_R\), the amplitude is reduced by
\begin{equation}
\exp(-|\omega_I|T_{\rm osc})
=
\exp\!\left(-\frac{\pi}{Q}\right).
\label{eq:q_decay_factor}
\end{equation}
For this range of \(Q\), the reduction factor is approximately
\(4\times10^{-2}\) to \(2\times10^{-1}\). Thus the corresponding KK excitations
decay within only a few oscillation cycles. This provides a quantitative
explanation for why they do not appear as narrow peaks in the real-axis
relative-probability and transmission spectra: they are broad, short-lived
dissipative modes rather than long-lived quasi-localized resonances.

\begin{table}[tbp]
\centering
\caption{\label{tab:quality_factors}
Quality factors \(Q=\omega_R/(2|\omega_I|)\) computed from the fundamental
QNFs in table~\ref{tab:qnm_comprehensive}. The values are of order unity,
indicating that the corresponding modes are strongly damped and short-lived.}
\vspace{0.2cm}
\resizebox{\textwidth}{!}{ 
\begin{tabular}{ccccccc}
\hline\hline
$B$ & $\alpha$ & Frequency-Domain & P \& Odd-Wave & DP \& Odd-Wave & P \& Even-Wave & DP \& Even-Wave \\
\hline
\multirow{4}{*}{1} 
& $0$        & $0.947$ & $0.985$ & $0.985$ & $0.969$ & $0.936$ \\
& $\alpha_1$ & $1.167$ & $1.192$ & $1.194$ & $1.188$ & $1.181$ \\
& $\alpha_2$ & $1.191$ & $1.198$ & $1.199$ & $1.210$ & $1.195$ \\
& $\alpha_s$ & $1.054$ & $1.090$ & $1.092$ & $1.081$ & $1.066$ \\
\hline
\multirow{4}{*}{2} 
& $0$        & $1.197$ & $1.150$ & $1.147$ & $1.164$ & $1.172$ \\
& $\alpha_1$ & --      & $1.388$ & $1.386$ & $1.362$ & $1.344$ \\
& $\alpha_2$ & $1.851$ & $1.777$ & $1.772$ & $1.795$ & $1.795$ \\
& $\alpha_s$ & $1.275$ & $1.281$ & $1.277$ & $1.255$ & $1.237$ \\
\hline
\multirow{4}{*}{4} 
& $0$        & $1.383$ & $1.450$ & $1.470$ & $1.558$ & $1.570$ \\
& $\alpha_1$ & --      & $1.452$ & $1.449$ & $1.485$ & $1.505$ \\
& $\alpha_2$ & --      & --      & --      & --      & --      \\
& $\alpha_s$ & --      & $1.367$ & $1.365$ & $1.415$ & $1.444$ \\
\hline\hline
\end{tabular}
}
\end{table}

\FloatBarrier

\section{Conclusion and discussion}
\label{sec:conclusion}

In this work, we investigated the tensor-sector spectroscopy of a thick brane in
\(f(R)=R+\alpha R^2\) gravity supported by two scalar fields. The two-scalar
sector provides an analytical realization of a warped thick-brane background
with tunable internal energy-density structure. In particular, by varying the
thickness parameter \(B\) and the curvature parameter \(\alpha\), the background
can interpolate between a single-peak energy-density profile and a
Bloch-brane-like split configuration. This construction allows us to examine how
the internal structure of the matter sector and the modified-gravity coupling
enter the tensor perturbation spectrum.

A central result of our analysis is that the Bloch-type internal structure does
not by itself generate narrow tensor resonances in the ghost-free region. In the
smooth branch where \(f_R>0\), the tensor effective potential is regular and
asymptotically vanishes, giving a gapless continuum of massive KK modes. The
relative-probability method shows no sharp real-axis peaks for the representative
ghost-free configurations. We further supported this conclusion by computing the
real-axis scattering phase shifts and the corresponding transmission
coefficient. The transmission spectrum increases smoothly with the KK mass and
does not exhibit narrow resonant peaks. Thus, the absence of peaks is not merely
a peculiarity of the relative-probability diagnostic; it reflects the fact that
the smooth ghost-free potentials are too transparent to support long-lived
quasi-localized tensor modes on the real mass axis.

The situation is qualitatively different in the singular branch. For sufficiently
large \(B\), when \(\alpha_k\leq\alpha\leq\alpha_2\), the effective coupling
\(f_R\) crosses zero and the tensor effective potential develops singular
structures through inverse powers of \(f_R\). In this branch, sharp peaks appear
in the relative-probability spectrum, and the corresponding wave functions are
strongly enhanced near the brane region. These peaks are therefore associated
with the singular \(f_R=0\) boundary rather than with the matter-sector
energy-density splitting alone. Since this branch violates the ghost-free
condition, the sharp real-axis peaks should be interpreted as
singular-boundary quasi-localization signals, not as ghost-free resonances of the
smooth brane background.

The absence of narrow real-axis peaks in the ghost-free branch does not imply
that the massive KK spectrum is physically featureless. Instead, it means that
the relevant spectral information is carried by broad, dissipative poles in the
complex frequency plane. We therefore extended the analysis to QNMs. The real
part of a QNF determines the oscillation scale of the KK ringdown, while the
imaginary part measures the leakage rate of the perturbation energy into the
extra dimension. All extracted tensor QNMs in the ghost-free region have
\(\omega_I<0\), consistently describing damped perturbations. The quality factors
computed from the fundamental modes are of order unity,
\(Q=\omega_R/(2|\omega_I|)\simeq 0.9-1.9\), showing that these modes decay within
only a few oscillation periods. This quantitatively explains why they do not
appear as narrow peaks in either the relative-probability spectrum or the
transmission spectrum: they are short-lived dissipative KK excitations rather
than long-lived real-axis resonances.

Methodologically, we combined frequency-domain and time-domain approaches. The
AIM provides direct access to the complex QNFs in cases where the potential is
suitable for the method, while the time-domain evolution offers a flexible
numerical approach for more general values of \(B\). The supersymmetric partner
potential was used as a practical tool to suppress the non-dissipative graviton
zero-mode contribution in the waveform and to isolate the massive-sector
ringdown more cleanly. The agreement among different time-domain strategies and
the available frequency-domain benchmarks indicates that the extracted QNFs are
stable features of the tensor perturbation problem rather than artifacts of a
specific initial wave packet or extraction method.

The physical picture that emerges is therefore a spectral classification of the
two-scalar \(f(R)\) thick brane. The ghost-free branch is characterized
by smooth potentials, no narrow real-axis tensor resonances, and short-lived
dissipative QNMs. The singular branch, by contrast, can produce sharp
quasi-localization peaks, but these are tied to the loss of the ghost-free
condition through \(f_R=0\). In this sense, the tensor spectrum acts as a
diagnostic of whether a given spectral structure originates from a regular
brane geometry or from a singular modified-gravity boundary.

Several extensions are worth pursuing. First, the present work focused on the
tensor sector; a full treatment of scalar and vector perturbations would be
needed to establish the complete perturbative stability of the model. Second,
it would be useful to follow the motion of QNM poles as \(\alpha\) approaches
the \(f_R=0\) boundary from the ghost-free side, in order to clarify how broad
dissipative modes are related to the sharp real-axis structures of the singular
branch. Finally, thick branes with de Sitter or anti-de Sitter induced metrics,
finite extra dimensions, or multi-brane configurations may possess different
boundary conditions and could lead to qualitatively distinct QNM spectra,
including echoes or more sharply defined quasi-bound structures.

\acknowledgments

The authors are grateful to Prof. Yu-Xiao Liu and  Yun-Peng E at Lanzhou
University for useful discussions on gravitational resonances in the singular
branch and on numerical methods for extracting quasinormal modes. Their
suggestions helped improve the spectral analysis presented in this work.

This work is supported by the National Natural Science Foundation of China
(Grants No.~11305119), the Natural Science Foundation of Shaanxi Province
(No.~2022JQ-037), and the 111 Project (B17035).

\end{document}